\documentstyle[preprint,aps]{revtex}
\renewcommand{\baselinestretch}{1.2}

\begin{document}
\title{Semi-analytical Proof of 
Abelian Dominance  \\ on Confinement in the Maximally Abelian Gauge}
\author{H. Ichie\footnote{E-mail: ichie@th.phys.titech.ac.jp}}
\address{
Department of Physics, Tokyo Institute of Technology \\
Meguro, Tokyo 152-8551, Japan
}
\author{H. Suganuma\footnote{E-mail: suganuma@rcnp.osaka-u.ac.jp}}
\address{
Research Center for Nuclear Physics (RCNP), Osaka University \\
Ibaraki, Osaka 567-0047, Japan }
\maketitle
\begin{abstract} 
We study abelian dominance for confinement
in terms of the local gluon properties
in the maximally abelian (MA) gauge
in a semi-analytical manner with the help of the lattice QCD.
The global Weyl symmetry persistently
remains as the relic of SU($N_c$) in the MA gauge,
and provides the ambiguity on the electric and magnetic charges.
We derive the criterion on the SU($N_c$)-gauge invariance
in terms of the residual symmetry in the abelian gauge.
In the lattice QCD,
we find microscopic abelian dominance on the link variable
for the whole region of $\beta$ in the MA gauge.
The off-diagonal angle variable, which is not constrained
by the MA-gauge fixing condition,
tends to be random besides the residual gauge degrees of freedom.
Within the random-variable approximation for
the off-diagonal angle variable,
we prove that off-diagonal gluon contribution to the Wilson loop
obeys the perimeter law in the MA gauge,
and show exact abelian dominance for the string tension,
although small deviation is brought by the finite size effect of
the Wilson loop in the actual lattice QCD simulation.
\end{abstract}
\newpage

\section{Introduction}
Quantum Chromodynamics (QCD) is the fundamental theory of the strong 
interaction\cite{cheng,kerson}.
Due to the asymptotic freedom, 
the gauge-coupling constant of QCD becomes small 
in the high-energy region
and the perturbative QCD provides a direct and systematic description 
of the QCD system in terms of quarks and gluons. 
On the other hand, in the low-energy region, 
the strong gauge-coupling nature of QCD leads to 
nonperturbative features like color confinement, 
dynamical chiral-symmetry breaking\cite{NJL,higashijima,miransky},
and nontrivial topological effect by instantons\cite{belavin,diakonov,shuryak}, 
and it is impossible to understand them directly from quarks and gluons 
in a perturbative manner. 
Instead of quarks and gluons, 
some collective or composite modes may be relevant degrees of freedom 
for the nonperturbative description 
in the infrared region of QCD. 
As for chiral dynamics,
the pion and the sigma meson play the important role for the low-energy 
QCD, and they are included in the QCD effective theories like the (non-) linear 
sigma model, 
the chiral bag model
\cite{thomasA,hosaka}
and the Nambu-Jona-Lasinio model
\cite{NJL,kunihiro},
where these mesons are
described as composite modes of quarks and anti-quarks.  
Here, the pion is considered to be the Nambu-Goldstone boson relating to 
spontaneous chiral-symmetry breaking and obeys the 
low-energy theorem\cite{bando} and the current algebra\cite{cheng}.
On the other hand, confinement is essentially described by  dynamics of
gluons rather than quarks.
Hence, it is desired to extract the relevant collective mode 
from gluons for confinement phenomena.

In 1970's, Nambu, 't Hooft and Mandelstam proposed 
an interesting idea that quark confinement can be interpreted 
using the dual version of the 
superconductivity\cite{nambu,thoa,mandelstam}.
In the ordinary superconductor, 
Cooper-pair condensation leads to the Meissner effect, 
and the magnetic flux is excluded or squeezed like a 
quasi-one-dimensional tube as the Abrikosov vortex\cite{abrikosov}, 
where the magnetic flux is quantized topologically. 
On the other hand, 
from the Regge trajectory\cite{greiner} of hadrons and the lattice 
QCD simulation\cite{rothe,creutz}, 
the confinement force between the color-electric charge is 
characterized by the universal physical quantity of the string tension, 
and is brought by one-dimensional squeezing of the 
color-electric flux\cite{haymaker} in the QCD vacuum. 
Hence, the QCD vacuum would be regarded as the dual version 
of the superconductor based on above similarities 
on the low-dimensionalization of the quantized flux between charges. 
In this dual-superconductor picture for the QCD vacuum, 
the squeezing of the color-electric flux between quarks 
is realized by the dual Meissner effect
as the result of condensation of color-magnetic monopoles, which is the 
dual version of electric-charge condensation. 
Monopole condensation and its relevant role for confinement were 
analytically pointed out by Seiberg and Witten very recently in the 
$N=2$ supersymmetric version of QCD\cite{seiberg}.

However, there are two following large gaps between QCD and the dual 
superconductor picture.
\begin{enumerate}
\item This picture is based on the abelian gauge theory subject to the 
Maxwell-type equations, where electro-magnetic duality is manifest, 
while QCD is a nonabelian gauge theory.
\item The dual-superconductor scenario  requires condensation of 
(color-)magnetic 
monopoles as the key concept, while QCD does not have such a monopole as 
the elementary degrees of freedom.
\end{enumerate}
\noindent As the connection between QCD and the dual superconductor scenario, 
't~Hooft 
proposed the concept of the abelian gauge fixing\cite{thooft}, the partial
gauge fixing which only remains abelian gauge degrees of freedom in QCD.
By definition, the abelian gauge fixing reduces QCD into an abelian gauge
theory, where the off-diagonal element of the gluon field behaves as a 
charged matter field
and provides a color-electric current in terms of the residual abelian gauge 
symmetry.
As a remarkable fact in the abelian gauge, color-magnetic monopoles
appear as topological objects corresponding to 
the nontrivial homotopy group $\Pi_2( {\rm SU}(N_c)/{\rm U(1)}^{ N_c-1}) =
{\bf Z}^{ N_c-1}_\infty$.
Thus, by the abelian gauge fixing, QCD is reduced into an abelian
gauge theory including both the electric current $j_\mu$ and 
the magnetic current $k_\mu$, which is expected to provide the theoretical 
basis of the monopole-condensation scheme for the confinement 
mechanism.

As for the irrelevance of  the off-diagonal gluon gluons,
Ezawa and Iwazaki assumed abelian dominance that 
the only abelian gauge fields with monopoles 
would be essential for the description of nonperturbative 
phenomena in the low-energy region of QCD, 
and showed a possibility of monopole condensation in the infrared scale by 
investigating ``energy-entropy balance'' on the monopole 
current\cite{ezawa,thomas,ichie2,ichie5} 
in a similar way to 
the Kosterliz-Thouless transition in the 1+2 dimensional superconductivity
\cite{kosterlitz}.
Ezawa and Iwazaki formulated the dual London theory as an infrared 
effective theory of QCD, 
and later it is reformulated as the
dual Ginzburg-Landau 
theory\cite{maedan,suganuma,sasaki,umisedo,ichie1,kondo,atanaka}.

Furthermore, abelian dominance\cite{yotsuyanagi,hioki,bali,miyamura,woloshyn} and 
monopole 
condensation\cite{kronfeld,schierholz,giacomo}
have been investigated using 
the lattice QCD simulation in the maximally abelian (MA)
gauge\cite{kronfeld,schierholz,giacomo,kondo2}.
The MA gauge is the abelian gauge where the diagonal component of the gluon 
is maximized by the gauge transformation.
In the MA gauge, physical information of the gauge configuration is 
concentrated into the diagonal component as much as possible.
The lattice QCD studies indicate {\it abelian dominance} 
that the string tension\cite{yotsuyanagi,hioki,bali} and 
the chiral condensate\cite{miyamura,woloshyn} are almost described only 
by abelian variables 
in the MA gauge. 
Moreover, 
{\it monopole dominance} is also observed in the lattice QCD simulation
in the MA gauge: 
only the monopole part in the abelian variable 
contributes to the nonperturbative QCD\cite{bali,miyamura}.
Thus, the lattice QCD phenomenology suggests the dominant role of 
abelian variables including monopoles 
for the nonperturbative QCD in the MA gauge\cite{giacomo2,poly}.

In this paper, we aim to understand the origin of abelian dominance
for confinement in terms of the local gluon properties in the MA gauge,
and study the relation between macroscopic abelian dominance and
microscopic abelian dominance in a semi-analytical manner
with the help of the lattice QCD simulation.
In Section 2, we study the residual symmetry
and the gauge invariance condition for operators
in the 't~Hooft abelian gauge,
paying the attention to the global Weyl symmetry.
In Section 3, we investigate the MA gauge 
both in the lattice and in the continuum theories
in terms of the gauge connection.
In Section 4,
we introduce the abelian projection rate
as the overlapping factor between SU(2) and abelian link variables,
and study microscopic abelian dominance in the MA gauge
in the lattice formalism.
In Section 5, we study the contribution of off-diagonal gluons
to the Wilson loop in the MA gauge,
and prove abelian dominance for the string tension
in a semi-analytical manner.
Section 6 is devoted to summary and concluding remarks.

\section{Residual Symmetry and Gauge Invariance Condition in the 
Abelian Gauge}

The dual superconductor picture for confinement phenomena 
is based on the abelian gauge theory including monopoles, 
and the 't Hooft abelian gauge fixing\cite{thooft} 
is the key concept for the connection from QCD 
to such an abelian gauge theory.
In this section, we investigate the abelian gauge fixing in QCD
in terms of the residual gauge symmetry.

The abelian gauge fixing,
the partial gauge fixing which remains the abelian gauge symmetry, 
is realized by the diagonalization of a 
suitable SU($N_c$)-gauge dependent variable 
as $\Phi[A_\mu(x)] \in su(N_c) $
by the SU($N_c$) gauge transformation.
In the abelian gauge, $\Phi[A_\mu(x)]$ plays the 
similar role of the Higgs field, and 
can be regarded as the composite Higgs field. 

For an hermite operator $\Phi[A_\mu(x)]$ which 
obeys the adjoint transformation,
$\Phi(x)$ is transformed as
\begin{eqnarray}
\Phi(x) = \Phi^a T^a 
\rightarrow \Phi^\Omega(x) & =  & \Omega(x) \Phi(x) \Omega^{\dagger}(x) 
\equiv \vec H \cdot \vec \Phi_{diag}(x)  \nonumber \\ 
&  = & {\rm diag}(\lambda^1(x), \cdots, \lambda^{N_c}(x)),
\end{eqnarray}
using a suitable gauge function 
$\Omega(x) = {\rm exp} \{ i \xi^a (x) T^a \} \in$ SU($N_{c}$).
Here, each diagonal component $\lambda^{i}$ 
($i$=1,$\cdots$,$N_c$) is to be real for the 
hermite operator $\Phi[A_\mu(x)]$.
In the abelian gauge, the SU($N_{c}$) gauge symmetry is reduced into 
the U(1)$^{N_{c}-1}$ gauge symmetry 
corresponding to the gauge-fixing ambiguity. 
The operator $\Phi(x)$ is diagonalized to 
$\vec H \cdot \vec \Phi_{diag}(x)$ also 
by the gauge function $\Omega^\omega (x) \equiv \omega(x)\Omega(x)$ with 
$\omega(x) = {\rm exp}(- i \, \vec H \cdot  \vec \varphi(x)  ) \in$
U(1)$^{N_{c}-1}$, 
\begin{eqnarray}
\Phi(x) \rightarrow  \Omega^\omega(x) \Phi(x)
\Omega^{\omega\dagger}(x) = \omega(x) \vec  H \cdot \vec \Phi_{diag}(x)
\omega^{\dagger}(x) = \vec H \cdot  \vec \Phi_{diag}(x),
\end{eqnarray}
and therefore U(1)$^{N_{c}-1}$ abelian gauge symmetry remains 
in the abelian gauge.

In the abelian gauge, there also remains the global Weyl symmetry as a 
``relic'' of the nonabelian theory \cite{suganuma4,suganuma2}.
Here, the Weyl symmetry corresponds to the subgroup of SU($N_{\rm c}$)
relating to the permutation of the basis in the fundamental representation.
Then, the Weyl group is expressed as 
the permutation group {\bf P}$_{N_c}$ including ${}_{N_c} C _2$ elements.
For simplicity, let us consider the $N_c=2$ case.  
For SU(2) QCD, the Weyl symmetry corresponds to the 
interchange of the SU(2)-quark color, 
$| + \rangle \equiv $
$({}^1_0)$ and $| - \rangle \equiv $ $({}^0_1)$, in the fundamental 
representation.
The global Weyl transformation is expressed by the global gauge function,
\begin{eqnarray}
W  & =  & 
e^{ i \{  \frac{\tau_1}{2} \cos \alpha + \frac{\tau_2}{2}\sin \alpha 
 \} \pi }  
=
i \{ \tau_1 \cos \alpha + \tau_2 \sin \alpha \}  \nonumber \\
 & = & i \left( 
{\matrix{
0  & e^{-i \alpha}  \cr
e^{i \alpha} & 0 \cr
}}
\right)
\hspace{0.2cm} \in  {\bf P}_2 \subset \mbox{SU}(2)
\label{eq:weyl}
\end{eqnarray}
with an arbitrary constant $\alpha \in {\bf R}$. 
By the global Weyl transformation $W$, 
the SU(2)-quark color is interchanged 
as $W | + \rangle = i e^{i \alpha} | - \rangle $ and 
$W | - \rangle = i e^{-i \alpha} | + \rangle $ except 
for the global phase factor.
This global Weyl symmetry remains in the abelian gauge, 
because the operator $\Phi(x)$ is also diagonalized by using
$\Omega^W(x) \equiv W \Omega(x)$, 
\begin{eqnarray}
\Phi(x) \rightarrow  \Omega^{W}(x) \Phi(x)
\Omega^{W{\dagger}}(x) = W \Phi_{diag}(x) \frac{\tau_3}{2} W ^{\dagger}
= - \Phi_{diag}(x) \frac{\tau_3}{2},
\end{eqnarray}
Here, the sign of $\Phi_{diag}(x)$, or the order of 
the diagonal component $\lambda^{i}(x)$, is globally changed 
by the Weyl transformation. 
It is noted that 
the sign of the U(1)$_3$ gauge field 
${\cal A}_\mu \equiv A^3_\mu \frac{\tau_3}{2}$ is globally changed
under the Weyl transformation,
\begin{eqnarray}
{\cal A}_\mu \rightarrow {\cal A}_\mu^{W} = W  A^3_\mu \frac{\tau_3}{2} W^\dagger = 
- A^3_\mu \frac{\tau_3}{2} = -{\cal A}_\mu.
\end{eqnarray}
Therefore, all the sign of the abelian field strength,
electric and  magnetic charges are also globally changed:
\begin{eqnarray}
{\cal F}_{\mu\nu}  & \equiv & F_{\mu\nu}\frac{\tau_3}{2}  \rightarrow   
{\cal F}_{\mu\nu}^{W} = W  {\cal F}_{\mu\nu}  W^\dagger 
= - {\cal F}_{\mu\nu},  \nonumber \\
j_\mu & \equiv & \partial^\alpha {\cal F}_{\alpha\mu}   
\rightarrow   j_\mu^{W}=-j_\mu, \nonumber \\
k_\mu & \equiv & \partial^\alpha {}^*{\cal F}_{\alpha\mu}   
\rightarrow   k_\mu^{W}=-k_\mu.
\end{eqnarray}
In the abelian gauge, the absolute signs of the electric 
and the magnetic charges are settled, only when the Weyl symmetry 
is fixed by the additional condition.
When $\Phi[A_{\mu}(x)]$ obeys the adjoint-type gauge transformation 
like the nonabelian Higgs field, the global Weyl symmetry can be 
easily fixed by imposing the additional gauge-fixing condition 
as $\Phi_{\it diag} (x) \ge 0$ for SU(2), or 
the ordering condition of the diagonal components $\lambda^{i} $ 
in $\vec H \cdot \vec \Phi_{\it diag}$ 
as $\lambda^{1} \ge..\ge \lambda^{N_c}$ for the SU($N_{c}$) case. 
As for the appearance of monopoles in the abelian gauge, 
the global Weyl symmetry ${\bf P}_{N_c}$ is not relevant, because 
the nontriviality of the homotopy group is 
not affected by the global Weyl symmetry. 
However, the definition of the magnetic monopole charge, 
which is expressed by the nontrivial dual root 
of ${\rm SU}(N_c)_{\rm dual}$ \cite{ezawa}, 
is globally changed by the Weyl transformation.

Now, we consider the abelian gauge fixing 
in terms of the coset space of the fixed gauge symmetry.
The abelian gauge fixing is a sort of the partial 
gauge fixing which reduces the gauge group 
$G \equiv $SU($N_{c})_{\rm local}$ of the system 
into its subgroup $H \equiv 
$U(1) $^{N_{c}-1}_{\rm local} (\times {\rm P}_{N_c}^{\rm global})$ 
including the maximally torus subgroup of G. 
In other words, the abelian gauge fixing freezes the gauge symmetry 
relating to the coset space $G/H$, 
and hence the representative gauge function $\Omega$ which brings 
the abelian gauge belongs the coset space $G/H$: $\Omega \in G/H$. 
In fact, $\Omega \in G/H$ is uniquely determined 
without the ambiguity on the residual symmetry $H$, if 
the additional condition on $H$ is imposed for $\Omega$.

However, such a partial gauge fixing makes the total gauge 
invariance unclear. Here, let  us consider 
the SU($N_c$) gauge-invariance condition on the 
operator defined in the abelian gauge\cite{suganuma4}.
To begin with, 
we investigate the gauge-transformation property of the 
gauge function $\Omega \in G/H$ which brings the abelian 
gauge (See. Fig.1). 
For simplicity, the operator $\Phi $ to be diagonalized is assumed to 
obey 
the adjoint gauge transformation as $\Phi \to \Phi ^g=g\Phi 
g^\dagger$ with $^\forall g\in G$.
After the gauge transformation by $^\forall g\in G$, 
$\Omega ^g\in {G / H}$ is defined so as to diagonalize $\Phi ^g$ as
$\Omega ^g\Phi ^g(\Omega ^g)^\dagger 
=\Phi 
_{diag}$, 
and hence the gauge function 
$\Omega ^g\in {G / H}$ which realizes the abelian gauge is 
transformed as
\begin{eqnarray}
\Omega \rightarrow \Omega ^g=h[g]\Omega g^\dagger
\label{eq:function}
\end{eqnarray}
under arbitrary SU($N_c$) gauge transformation by $g\in G$.
Here, $h[g] \in H$ is chosen so as to make $\Omega^g$ belong 
{\it G/H}, i.e., $\Omega^g \in G/H$. 
(If the additional condition on $H$ is imposed 
to specify $\Omega \in G/H$, $\Omega g^\dagger$ 
does not satisfy it in general.) 
This is similar to the argument on the hidden local 
symmetry\cite{bando} in the nonlinear representation.
In general, the gauge function $\Omega \in G/H$ is transformed 
nonlinearly by the gauge function $g$ due to $h[g]\in H$.
Thus, the gauge-transformation property 
on the gauge function $\Omega \in G/H$ 
becomes nontrivial in the partial gauge fixing.

Owing to the nontrivial transformation (\ref{eq:function}) of $\Omega 
\in$$G/H$, any operator $O^{\Omega}$ defined in the abelian gauge is 
found to be transformed as $O^{\Omega} \to (O^{\Omega})^{h[g]}$ by
the SU($N_c$) gauge transformation of ${}^\forall g \in G$.
We demonstrate this for the gluon field $A_\mu ^\Omega \equiv 
\Omega 
(A_\mu +{i \over e}\partial _\mu )\Omega ^\dagger$ in the  abelian 
gauge.
By the gauge transformation of $^\forall g\in G$, $A_\mu ^\Omega $ 
is 
transformed as
\begin{eqnarray}
A_\mu ^\Omega \to (A_\mu ^g)^{\Omega ^g}=A_\mu ^{\Omega ^gg}=A_\mu 
^{h[g]\Omega } = (A_{\mu}^\Omega)^{h[g]}
 =h[g](A_\mu ^\Omega +{i \over e}\partial _\mu )h^\dagger[g].
\end{eqnarray}
Here, we have used 
\begin{eqnarray}
(A_\mu ^{g_{1}})^{g_{2}}
& = &  g_{2}( A_\mu ^{g_{1}} +{i \over e}\partial _\mu ) g_{2}^\dagger
=
( g_{2} g_{1}) (A_\mu  +{i \over e}\partial _\mu ) ( g_{2} 
g_{1})^\dagger = (A_\mu)^{ g_{2} g_{1}} 
\end{eqnarray} 
for the successive gauge transformation by 
$g_{1}$, $g_{2} \in {\rm SU}(N_c)$.
Similarly, the operator $O^\Omega $ defined in the abelian gauge is 
transformed by
$^\forall g\in G$ as
\begin{eqnarray}
O^\Omega \to (O^g)^{\Omega ^g} &  = & \Omega ^gO^g\Omega 
^{g\dagger} = h[g]\Omega g^\dagger  \cdot gOg^\dagger \cdot g\Omega 
^\dagger h^\dagger[g]  \nonumber \\
 & = & h[g]\Omega O\Omega ^\dagger h^\dagger[g]=h[g]O^\Omega 
h^\dagger[g]=(O^\Omega )^{h[g]},
\end{eqnarray}
as shown in Fig.1.
Here, $O$ is assumed to obey the adjoint transformation as 
$O^g = gOg^\dagger$ 
for simplicity.

Thus, arbitrary SU($N_c$) gauge transformation by $g\in G$ is mapped into 
the partial gauge transformation by $h[g]\in H$ for the operator 
$O^\Omega $ defined in the abelian gauge, and $O^\Omega $ transforms 
nonlinearly as $O^\Omega \to (O^\Omega )^{h[g]}$ by the SU($N_c$) gauge 
transformation $g$.
If the operator $O^\Omega $ is $H$-invariant, one gets 
$(O^\Omega)^{h[g]}=O^\Omega $ for any $h[g] \in H$, 
and hence $O^\Omega$ is also $G$-invariant 
or total SU($N_c$) gauge invariant, 
because $O^\Omega $ is transformed into $(O^\Omega)^{h[g]}=O^\Omega$ 
by $^\forall g\in G$. 
Thus, we find a useful criterion on the SU($N_c$) gauge 
invariance of the operator defined in the abelian gauge \cite{suganuma4}:
If the operator $O^\Omega$ defined in the abelian gauge is $H$-invariant,
$O^\Omega$ is 
also invariant under the whole gauge transformation of $G$.

Here, let us consider the application of this criterion 
to the effective theory of QCD in the abelian gauge, 
the dual Ginzburg-Landau (DGL) theory\cite{maedan,suganuma}. 
In the DGL theory, 
the local U(1)$^{N_c-1}$ and the global Weyl symmetries remain, 
and the dual gauge field $B_{\mu}$ and the monopole field 
$\chi_\alpha$ [$\alpha$=1, $\cdots$, $\frac12 N_{c}(N_{c}-1)$]
are the relevant modes for infrared physics. 
Although $B_{\mu}$ is invariant under the local transformation of 
${\rm U(1)}^{N_c-1} \subset {\rm SU}(N_c)$, 
$B_{\mu} \equiv \vec B_{\mu} \cdot \vec H$ is variant 
under the global Weyl transformation, 
and therefore $B_{\mu}$ is SU($N_c$)-gauge dependent object 
and does not appear in the real world alone.
As for the monopole field, 
there exists one Weyl-invariant combination of 
the monopole field fluctuation, 
$\tilde \chi \equiv \sum_{\alpha}  \tilde \chi_{\alpha}$\cite{ichie1},
which is also {\rm U(1)}$^{N_c-1}$-invariant. 
Therefore, the monopole fluctuation $\tilde \chi$ 
is completely residual-gauge invariant in the abelian gauge, 
so that $\tilde \chi$ is SU($N_c$)-gauge invariant 
and is expected to appear as a scalar glueball with $J^{PC}=0^{++}$, 
like the Higgs particle in the standard model.

\section{Maximally Abelian (MA) Gauge in the Connection Formalism}

The abelian gauge has some arbitrariness
corresponding to the choice of the operator $\Phi$ to be diagonalized. 
As the typical abelian gauge, 
the maximally abelian (MA) gauge, the Polyakov gauge and the F12 gauge  
have been tested on the dual superconductor scenario for the 
nonperturbative QCD.
Recent lattice QCD studies show that 
infrared phenomena such as confinement properties and chiral symmetry
breaking are almost reproduced in the MA 
gauge\cite{yotsuyanagi,hioki,bali,miyamura,woloshyn,kronfeld,schierholz}. 
In the SU(2) lattice formalism, the MA gauge is defined so as to maximize
\begin{eqnarray}
R_{\rm MA}[U_\mu]  & \equiv &
\sum_{s,\mu} {\rm tr} \{  U_\mu(s) \tau_3 U^{\dagger}_\mu(s) \tau_3 \}  
\nonumber \\
   & = & 
2\sum_{s,\mu}\{
U^0_\mu(s)^2+U^3_\mu(s)^2-U^1_\mu(s)^2-U^2_\mu(s)^2 \} \nonumber \\
   & = & 
2\sum_{s,\mu} \left[
1 - 2\{ U^1_\mu(s)^2+U^2_\mu(s)^2 \} \right]
\end{eqnarray}
by the SU(2) gauge transformation.
Here, the link variable 
$U_{\mu}(s) \equiv U_{\mu}^0(s) + i \tau^a U_{\mu}^a (s) \in$ SU(2) with
$U_{\mu}^0(s)$, $U_{\mu}^a(s) \in$ {\bf R} relates to the (continuum) 
gluon field $A_\mu \equiv A_\mu^a T^a \in$ $su$(2) as $U_{\mu}(s) = 
e^{iaeA_{\mu}(s)}$, where $e$ denotes the QCD gauge coupling and $a$ the 
lattice spacing.   
In the MA gauge, the absolute value of off-diagonal components, 
$U_\mu^1(s)$ and $U_\mu^2(s)$, are
forced to be small.
In  the continuum limit $a \rightarrow 0$,
the link variable reads 
$U_\mu(s) = e^{iaeA_\mu(s)} = 1+iae A_{\mu}(s) + O(a^2)$,
and hence the MA gauge is found to minimize the functional, 
\begin{eqnarray}
 R_{ch}[A_\mu] \equiv \frac12 e^2 \int d^4 x \{ A_\mu^1(x)^2 + 
A_\mu^2(x)^2 \}
= e^2 \int d^4 x A_\mu^+(x) A_\mu^-(x),
\end{eqnarray}
with $A_\mu^\pm (x) \equiv {1 \over {\sqrt 2}}
  \{ A_\mu^1(x) \pm 
i A_\mu^2(x) \}$.
Thus, in the MA gauge, the off-diagonal gluon component is globally forced 
to be small by the gauge transformation, and hence the QCD system 
is expected to be describable only by its diagonal part
approximately.

The MA gauge is a sort of the abelian gauge which diagonalizes the 
hermite operator
\begin{eqnarray}
\Phi[U_{\mu}(s)] \equiv \sum _{\mu,\pm } U_{\pm \mu}(s) \tau_3 U^{\dagger}_{\pm 
\mu}(s).
\end{eqnarray}
Here, we use the convenient notation $U_{-\mu}(s) \equiv
U^{\dagger}_\mu(s- \hat \mu)$ in this paper.
In the continuum limit, the condition of the MA gauge becomes 
$\displaystyle \sum_\mu$$(i\partial_\mu \pm e A^3_{\mu} ) A^{\pm }_\mu = 
0$.
This condition can be regarded as the maximal decoupling condition 
between the abelian gauge sector and the charged gluon sector.

In the MA gauge, $\Phi(s)$ is diagonalized as 
$\Phi_{\rm MA}(s) = \Phi_{diag}(s) \tau_{3}$ with  $\Phi_{diag}(s) \in$ 
{\bf R}, and there remain the local U(1)$_3$ symmetry and the  global Weyl 
symmetry
\cite{suganuma4}.
As a remarkable fact, $\Phi(s)$ does not obey the adjoint 
transformation  in the MA gauge, and  
the sign of $\Phi_{diag}(s)$ 
is not changed by the  Weyl transformation by $W $ in 
Eq.(\ref{eq:weyl}), 
\begin{eqnarray}
\lefteqn{\Phi_{\rm MA}(s) = \Phi_{diag}(s) \tau_3 } \nonumber \\ 
  & \rightarrow &   
\Phi_{\rm MA}^W (s)  =  
\sum _{\mu,\pm }  W  U_{\pm \mu}(s)  W ^{\dagger}
\tau_3  W  U^{\dagger}_{\pm \mu}(s)
 W^{\dagger} \nonumber \\
&  & = 
- \sum _{\mu,\pm } W  U_{\pm \mu}(s)   
\tau_3  U^{\dagger}_{\pm \mu}(s)
 W ^{\dagger}
= - W \Phi_{diag}(s) \tau_3 W^\dagger  = \Phi_{diag}(s) \tau_3. 
\end{eqnarray}
Thus, the Weyl symmetry is not fixed in the MA gauge by the simple 
ordering condition as $ \Phi_{diag} \ge 0$, unlike the adjoint case.  
We find the gauge invariance condition on the operator $O^\Omega$
defined in the MA gauge:
if $O^\Omega$ is invariant both under the local U$(1)^{N_c-1}$ 
gauge transformation and the global Weyl transformation,
$O^\Omega$ is also invariant under 
the SU($N_c$) gauge transformation.

In the continuum SU($N_c$) QCD, 
it is more fundamental and convenient to define the MA gauge fixing 
by way of the SU($N_c$)-covariant derivative operator 
${\hat D_\mu} \equiv {\hat \partial_\mu} + ie A_\mu $, 
where $\hat \partial_\mu$ is the derivative operator 
satisfying $[\hat \partial_\mu, f(x)]= \partial_\mu f(x)$. 
The MA gauge is defined so as to make 
SU($N_{c}$)-gauge connection $\hat D _{\mu}= \hat \partial_{\mu} + ie 
A_{\mu}^{a} T^{a}$ close to U(1)${}^{N_{c}-1}$-gauge connection 
$\hat D _{\mu}^{\rm Abel}= \hat \partial_{\mu} + ie \vec A_{\mu} \cdot 
\vec H$ by minimizing 
\begin{eqnarray}
 R_{\rm ch} \equiv \int d^4 x \; {\rm tr}[{\hat D_\mu}, \vec 
H]^\dagger
[{\hat D_\mu}, \vec H]
= e^{2} \int d^{4}x \; {\rm tr}[A_\mu, \vec H]^{\dagger}[A_\mu, \vec H] 
\nonumber \\
= e^{2} \int d^4 x \; \sum_{\alpha, \beta}A_\mu^{\alpha*}A_\mu^{\beta}
\vec \alpha \cdot \vec \beta {\rm tr} (E_\alpha^{\dagger} E_\beta)
= \frac{e^{2}}{2} \int d^{4}x \sum_{\alpha = 1}^{N_c(N_{c}-1)}
|A_{\mu}^{\alpha}|^{2}, 
\end{eqnarray}
which expresses the total amount of the off-diagonal gluon component. 
Here, we have used the Cartan decomposition,
$\displaystyle
A_{\mu} \equiv A_\mu^{a} T^{a} = \vec A_{\mu}  \cdot \vec H +
\sum_{\alpha = 1}^{N_c(N_{c}-1)} A_{\mu} ^{\alpha} E^{\alpha}$; 
$\vec H \equiv  (T_3, T_8, \cdots, T_{N_c^2-1})$ is the Cartan 
subalgebra, and $E^{\alpha}(\alpha=1,2,\cdots,N_{c}^{2}-N_c)$ 
denotes the raising or lowering operator.
In this definition with $\hat D _{\mu}$, the gauge transformation 
property of $ R_{\rm ch}$ becomes quite clear, 
because the SU($N_c$) covariant derivative ${\hat D_\mu}$
obeys the simple adjoint gauge transformation, 
${\hat D_\mu} \rightarrow 
\Omega {\hat D_\mu} \Omega^{\dagger}$,
with the SU($N_c$) gauge function $\Omega \in$ SU($N_c$).
By the SU($N_c$) gauge transformation, $R_{\rm ch}$ is transformed as
 \begin{eqnarray}
R_{\rm ch} \rightarrow R_{\rm ch}^{\Omega} =
 \int d^{4} x \; {\rm tr}
 \left( [\Omega \hat D_{\mu} \Omega^{\dagger}, \vec H]^{\dagger}
[\Omega \hat D_{\mu} \Omega^{\dagger}, \vec H] \right) \nonumber \\
=     \int d^{4} x \; {\rm tr} 
 \left( [\hat D_{\mu}, \Omega^{\dagger}\vec H \Omega]^{\dagger}
[\hat D_{\mu}, \Omega^{\dagger}\vec H \Omega]\right),
\end{eqnarray} 
and hence
the residual symmetry corresponding to the invariance of $R_{\rm ch}$ 
is found to be U(1)$^{N_{c}-1}_{\rm local} \times P^{N_{c}}_{\rm 
global} \subset $SU($N_{c})_{\rm local}$, where  $P^{N_{c}}_{\rm 
global}$ denotes the global Weyl group relating to the permutation of 
the $N_{c}$ basis in the fundamental representation.
In fact, one finds $\omega^{\dagger} \vec H \omega = \vec H$ for 
$\omega = e^{-i \vec \varphi(x) \cdot \vec H} \in$   
U(1)$^{N_{c}-1}_{\rm local}$, and the global Weyl transformation by 
$W \in$ $P^{N_{c}}_{\rm global}$ only exchanges the permutation of the 
nontrivial root $\vec \alpha_{j}$ and never changes $R_{\rm ch}$.
In the MA gauge, arbitrary gauge transformation by ${}^\forall \Omega \in 
G$ is to increase $R_{\rm ch}$ as $R_{\rm ch}^{\Omega} \ge R_{\rm ch}$.
Considering arbitrary infinitesimal gauge transformation $\Omega = 
e^{i \varepsilon} \simeq 1 + i \varepsilon $ with ${}^\forall \varepsilon 
\in$$su$($N_{c}$), one finds $\Omega^{\dagger} \vec H \Omega \simeq \vec 
H + i[\vec H, \varepsilon]$ and
\begin{eqnarray}
R^{\Omega}_{\rm ch} \simeq  R_{\rm ch} + 2i\int d^{4}x 
{\rm tr} \left( [\hat D_{\mu}, [\vec H, \varepsilon]] [\hat D_{\mu}, \vec H] 
\right)
\nonumber \\
= R_{\rm ch} + 2i\int d^{4}x 
{\rm tr}\left( \varepsilon
[\vec H, [\hat D_{\mu},[\hat D_{\mu},\vec H] ]] \right).
\end{eqnarray}
In the MA gauge, the extremum condition of $R^{\Omega}_{\rm ch}$ on 
${}^\forall \varepsilon \in $$su$($N_{c}$) provides 
\begin{eqnarray}
[\vec H, [\hat D_{\mu}, [\hat D_{\mu},\vec H] ]] = 0,
\end{eqnarray}
which leads to $\sum_{\mu}(i \partial_{\mu} \pm eA^{3}_{\mu}) 
A_{\mu}^{\pm} = 0$ for the $N_{c}$=2 case.
Thus, the operator $\Phi$ to be diagonalized in the MA gauge is found 
to be
\begin{eqnarray}
\Phi[A_{\mu}] = [\hat D_{\mu}, [\hat D_\mu, \vec H]] 
\end{eqnarray}
in the continuum theory.

\section{ Microscopic Abelian Dominance in the MA gauge}

In the abelian gauge, the diagonal and the off-diagonal gluons 
play different  roles in terms of the residual abelian gauge symmetry: 
the diagonal gluon behaves as the abelian gauge field,
while off-diagonal gluons behave as charged matter 
fields\cite{kronfeld}.
Under the U(1)$_3$ gauge transformation by   
$\omega = {\rm exp}(- i \varphi \frac{\tau_3}{2}) \in$ U(1)$_3$,
one finds
\begin{eqnarray}
A_\mu^3 & \rightarrow & (A_\mu^{\omega})^3 = A_\mu^3  + \frac{1}{e} 
\partial_\mu 
\varphi \\ 
A_\mu^{\pm} & \rightarrow & (A_\mu^{\omega})^\pm = A_\mu^{\pm} e^{ \pm 
i\varphi}
\end{eqnarray} 
with $A^\pm_\mu = \frac{1}{\sqrt{2}} (A^1_\mu \pm i A^2_\mu)$.
The abelian projection is simply defined as the 
replacement of the gluon field $A_\mu = A_\mu^a {\tau^a \over 2} \in $ 
$su(2)$ by the diagonal part ${\cal A}_\mu \equiv  A_\mu^3 {\tau^3 \over 2} 
\in u(1)_3 \subset su(2)$.
 
We call
``abelian dominance for an operator $\hat O[A_\mu]$'', when the
expectation value $\langle \hat O \rangle$ is almost unchanged by 
the  abelian projection $A_{\mu} \to {\cal A}_{\mu}$ as 
$\langle \hat O[A_\mu] \rangle \simeq \langle \hat O[{\cal A}_\mu] 
\rangle_{\rm A.G.}$, when 
$\langle$ $\rangle_{\rm A.G.}$ denotes the expectation value in the 
abelian gauge. Ordinary abelian dominance is observed for the long-distance
physics in the MA gauge, and this would be physically interpreted as the 
effective-mass generation of the off-diagonal gluon 
induced by the MA gauge fixing\cite{amemiya,suganuma1}.

In the lattice formalism, the SU(2) link-variable $U_\mu(s)$ can be 
factorized as 
\begin{eqnarray}
U_\mu(s) & = & 
M_\mu(s)  u_\mu(s)   \hspace{2cm} \in G \nonumber \\
M_\mu(s) & = & 
{\rm exp} \left(
i \{  \tau_1 \theta^1_\mu(s)+ \tau_2 \theta^2_\mu(s)  \} 
\right) \hspace{1cm} 
\in  G/H, \nonumber \\
u_\mu(s) & = & {\rm exp} \left( {i \tau^3 \theta^3_\mu (s) } \right) 
\hspace{3cm} \in  H
\label{eq:para}
\end{eqnarray}
with respect to the Cartan decomposition of $G = G/H$ $\times$ $H$ into 
$G/H=$SU(2)$ /$U(1)$_3$ 
and $H=$U(1)$_3$.
Here, the abelian link variable, 
\begin{eqnarray}
u_\mu(s) = e^{i \tau^3  \theta^3_\mu (s) } =
 \left( {\matrix{
e^{i\theta^3_\mu  (s)}  &          0
\cr
0                     &  e^{-i\theta^3_\mu (s)} \cr
}} \right)  \hspace{1cm}  \in {\rm U(1)}_3 \subset {\rm SU(2)}, 
\end{eqnarray}
plays the similar role as the SU(2)-link variable 
$U_\mu(s) \in {\rm SU(2)}$ 
in terms of the residual U(1)$_3$ gauge symmetry 
in the abelian gauge, and $\theta^3_\mu(s) \in (-\pi, \pi]$ 
corresponds to the diagonal component of the gluon 
in the continuum limit. 
On the other hand, the off-diagonal factor 
$M_\mu(s) \in $SU(2)/U(1)$_3$ is expressed as 
\begin{eqnarray}
 M_\mu(s)  & = & 
{\rm exp} \left(
{i \{  \tau_1 \theta^1_\mu(s)+ \tau_2 \theta^2_\mu(s)  \} }
\right)
\nonumber \\
& = & 
\left( 
{\matrix{
{\rm cos}{\theta_\mu(s)} & -{\rm sin}{\theta_\mu(s)}e^{-i\chi_\mu(s)} \cr
{\rm sin}{\theta_\mu(s)}e^{i\chi_\mu(s)} & {\rm cos}{\theta_\mu(s)} 
}}
 \right) 
\label{divided} 
 \\ & = &
\left( {\matrix{
\sqrt{1-|c_\mu(s)|^2} & -c_\mu^*(s) \cr
c_\mu (s) & \sqrt{1-|c_\mu(s)|^2}
}} \right)  \nonumber 
\end{eqnarray}
with  
$\theta_\mu (s) \equiv {\rm mod}_\frac{\pi}{2} \sqrt{ (\theta^1_\mu )^2 + 
(\theta^2_\mu)^2} \in [0, \frac{\pi}{2}]$ and $\chi_\mu (s) \in ( - \pi, \pi]$.
Near the continuum limit, the off-diagonal elements of $M_\mu(s)$ 
correspond to the off-diagonal gluon components. 
Under 
the residual U(1)$_3$ gauge transformation by 
$\omega(s) = e^{-i \varphi(s) \frac{\tau_3}{2}}\in$ U(1)$_3$, 
$u_\mu(s)$ and $M_\mu(s)$ are transformed as
\begin{eqnarray}
u_\mu(s) &\rightarrow & 
u^\omega _\mu(s) = \omega(s) u_\mu(s) \omega^\dagger(s+ \hat \mu 
)  \hspace{1cm}  \in H\\
M_\mu(s) &\rightarrow &
M^\omega_\mu(s) = \omega(s) M_\mu(s) \omega^\dagger(s) \hspace{1cm} \in G/H
\end{eqnarray}
so as to keep $M_\mu^\omega(s)$ belong $G/H$.
Accordingly, 
$\theta^3_\mu(s)$ and $c_\mu(s) \in {\bf C}$ are transformed as 
\begin{eqnarray}
\theta^3_\mu(s) &\rightarrow & 
\theta^{3 \omega}_\mu(s)=
{\rm mod}_{2\pi} [ \theta^3_\mu(s) + \{ \varphi(s+ \hat \mu)-\varphi(s) \} /2 ]
\\
c_\mu(s) &\rightarrow & 
c_\mu^{\omega}(s) = c_\mu(s)e^{i\varphi(s)}.
\end{eqnarray}
Thus, on the residual U(1)$_3$ gauge symmetry, 
$u_\mu(s)$ behaves as the U(1)$_3$ lattice gauge field, and 
$\theta^3_\mu(s)$ behaves as the U(1)$_3$ gauge field 
in the continuum limit. 
On the other hand, 
$M_\mu(s)$ and $c_\mu(s)$ behave as the charged matter field 
in terms of the residual U(1)$_3$ gauge symmetry, which is 
similar to the charged weak boson $W_\mu^{\pm}$ in the standard model.

In this parameterization
(\ref{eq:para}), there are two U(1)-structures embedded in SU(2) 
corresponding to $e^{i \theta^3_\mu }$ and $e^{ i  \tilde { \chi}_\mu}$. 
To clarify this structure, we reparametrize the SU(2) link variable as
\begin{eqnarray}
U_\mu(s) & = &
 \left( {\matrix{
{\rm cos}{\theta_\mu}e^{i \theta^3_\mu } &
 -{\rm sin}{\theta_\mu}e^{-i {\tilde \chi}_\mu} \cr
{\rm sin}{\theta_\mu}e^{i {\tilde \chi}_\mu} & 
{\rm cos}{\theta_\mu}e^{-i \theta^3_\mu }
 }} \right), 
\label{eq:repara}
\end{eqnarray}
or equivalently
\begin{eqnarray}
U^0_\mu & = & \cos \theta _\mu \cos \theta^3_\mu, \hspace{1cm} 
U^1_\mu  =  \sin \theta _\mu \sin \tilde \chi_\mu, \nonumber \\
U^3_\mu & = & \cos \theta _\mu \sin \theta^3_\mu,  \hspace{1cm} 
U^2_\mu  =  \sin \theta _\mu \cos \tilde \chi_\mu,
\end{eqnarray}
with $\tilde \chi_\mu \equiv \chi_\mu +\theta^3_\mu$.
The range of the angle variable can be redefined as 
$0 \le \theta_\mu \le \frac{\pi}{2}$
 and $-\pi < \theta^3_\mu,  \tilde \chi_\mu \le \pi$.
Here, $(U^0_\mu,U^1_\mu,U^2_\mu,U^3_\mu)$ forms an element of the 
3-dimensional 
hyper-sphere $S^3 \simeq $SU(2), because of
$(U_\mu^0)^2 +(U_\mu^1)^2 + (U_\mu^2)^2 + (U_\mu^3)^2  = 1$.
 For a fixed $\theta_\mu$, both 
$(U^0_\mu,U^3_\mu)$ and
$(U^1_\mu,U^2_\mu)$ form the two $S^1 \simeq $U(1) subgroups embedded in 
$S^3$ in
a symmetric manner.
From the parametrization in Eq.(\ref{eq:repara}), 
the SU(2) measure can be easily found as 
\begin{eqnarray}
\int d U_\mu & \equiv & 
\int d U_\mu^0 U_\mu^1 U_\mu^2 U_\mu^3  \,\,\, \delta \,(  \,\sum_{a = 0}^3  
\,( \, U^a_\mu )^2 -1 ) \nonumber \\
& = &\frac{1}{2 \pi^2} 
\int_{0}^{\frac{\pi}{2}} d \theta_\mu \sin  \theta_\mu \cos \theta_\mu 
\int_{-\pi}^{ \pi} d \tilde \chi_\mu 
\int_{-\pi}^{ \pi} d \theta^3_\mu.
\label{eq:measure}
\end{eqnarray}
In the lattice formalism,
the abelian projection is defined by replacing the SU(2) link variable 
$U_\mu(s) \in$ SU(2) by the abelian link variable $u_\mu(s) \in$ U(1)$_3$.

In the MA gauge, the off-diagonal gluon component is strongly suppressed, and the 
SU(2) link variable is expected to be U(1)$_3$-like as 
$U_\mu(s) \simeq u_\mu(s)$ in the relevant gauge configuration.
In the quantitative argument, this can be expected as 
$\langle U_{\mu}(s) u_{\mu}^{\dagger}(s) \rangle_{\rm {MA}} \simeq 1$,
where $\langle \: \rangle_{\rm {MA}}$ denotes the expectation value in 
the MA gauge.
In order to estimate the difference between $U_{\mu}(s)$ and 
$u_{\mu}(s)$,
we introduce the ``abelian projection rate'' $R_{\rm {Abel}}$ 
\cite{poulis,ichie6},
which is 
defined as the overlapping factor as 
\begin{eqnarray}
R_{\rm Abel} (s,\mu) \equiv \frac12 {\rm Re} \; {\rm tr} \{ U_{\mu}(s) 
u_{\mu}^{\dagger}(s) \} = \frac12 {\rm Re} \; {\rm tr} M_{\mu}(s) = \cos 
\theta_\mu(s).
\end{eqnarray}
This definition of $R_{\rm {Abel}}$ is inspired by the ordinary 
``distance'' between two 
matrices $A, B \in {\rm GL}(N,{\bf C})$ defined as
$d^{2}(A,B) \equiv \frac12 {\rm tr} \{ (A-B)^{\dagger} (A-B) 
\}$\cite{georgi}, 
which 
leads to 
$d^{2}(A,B) = 2-{\rm Re} \;{\rm tr}(AB^{\dagger})$ for $A,B \in$SU(2).
The similarity between $U_\mu(s)$ and $u_\mu(s)$ 
can be quantitatively measured 
in terms of the ``distance'' between them.
For instance, if $\cos \theta _\mu (s) =1$,
the SU(2) link variable becomes completely abelian as
\begin{eqnarray}
U_\mu(s) =
 \left( {\matrix{
e^{i\theta^3_\mu}  &          0
\cr
0                     &  e^{-i\theta^3_\mu} \cr
}} \right), \nonumber
\end{eqnarray}
while, if $\cos \theta _\mu(s) =0$, it becomes completely off-diagonal as 
\begin{eqnarray}
U_\mu(s)  =
 \left( {\matrix{
0 & -e^{-i \tilde \chi_\mu}  
\cr
e^{i\tilde \chi_\mu} & 0\cr
}} \right). \nonumber
\end{eqnarray}

We show
in Fig.2(a) the probability distribution  $P(R_{\rm 
Abel})$  
of the abelian projection rate $R_{\rm Abel} (s,\mu) \equiv \cos \theta_\mu(s)$
in the MA gauge. 
Here, $\langle  R_{\rm Abel} \rangle_{\beta=0}$  
in the strong coupling limit ($\beta = 0$)\cite{poulis,ichie6} is analytically 
calculable as  
\begin{eqnarray}
\langle  R_{\rm Abel} \rangle _{\beta=0}   & = & 
\langle \cos \theta_\mu (s) \rangle _{\beta=0} = 
\frac{  \int d U_\mu(s) \cos \theta_\mu(s)}
 {  \int d U_\mu(s)}   \nonumber \\
& = &
\frac{\int_{0}^{\frac{\pi}{2}} d \theta_\mu(s) 
\sin \theta_\mu (s) \cos^2 \theta_\mu(s) }
 { \int_{0}^{\frac{\pi}{2}} d \theta_\mu (s) 
\sin \theta_\mu(s) \cos \theta_\mu(s)  } =
\frac23,
\end{eqnarray}
using Eq.(\ref{eq:measure}).
In the MA gauge, $R_{\rm Abel}$ approaches to unity as shown in Fig.2(a).
The off-diagonal component of the SU(2) link variable 
is forced to be reduced.
As a typical example, one obtains   
$\langle  R_{\rm Abel} \rangle_{\rm MA} \simeq$  0.926  on $16^4$ lattice
with $\beta = 2.4$.
We show also the abelian projection rate 
$\langle  R_{\rm Abel} \rangle_{\rm MA}$ as the function of $\beta$ in 
Fig.2(b).
For larger $\beta$, 
$\langle \cos \theta_\mu(s) \rangle_{\rm MA}$ becomes slightly larger.
Without gauge fixing, the average  
$\langle  R_{\rm Abel} \rangle$ is found to be about $2 \over 3$ 
without dependence on $\beta $.
In the continuum limit in the MA gauge,
$U_\mu^1(s) $ and  $U_\mu^2(s)$ become at most
$O(a)$,  and  therefore  $\langle  R_{\rm Abel} \rangle_{\rm MA}$
approaches to unity as  
$\langle  R_{\rm Abel} \rangle_{\rm MA} = 1 + O(a^2)$
due to the trivial dominance 
of $U_\mu^0(s)$, which differs from abelian dominance in the physical 
sense.
The remarkable feature of the MA gauge is  the high abelian 
projection rate as $\langle  R_{\rm Abel} \rangle_{\rm MA} \simeq 1$ in the 
whole region of $\beta$. In fact, we find  
$\langle  R_{\rm Abel} \rangle_{\rm MA} \simeq 
0.88 $ even for the strong coupling limit $\beta=0$, where the 
original link variable $U_\mu$ is completely random.
Thus, abelian dominance for the link variable $U_{\mu}$ is observed at 
any scale in the MA gauge.

To understand the origin of the high abelian projection rate  
as $\langle  R_{\rm Abel} \rangle_{\rm MA} \simeq 1$, 
we estimate the lower bound of 
$\langle  R_{\rm Abel} \rangle_{\rm MA} $ in the MA gauge using the 
statistical 
consideration.
The MA gauge maximizes
\begin{eqnarray}
R_{\rm MA}[U_\mu]   \equiv 
\sum_{s,\mu} {\rm tr} \{  U_\mu(s) \tau_3 U^{\dagger}_\mu(s) \tau_3 \}  
={\rm tr}(\tau_3  \sum_{s,\mu} \hat \phi_\mu(s)),
\end{eqnarray}
where $\hat  \phi_\mu(s) \equiv 
U_{\mu}(s) \tau_3 U^{\dagger}_{\mu}(s)$
is an  $su$(2) element satisfying $\hat \phi^2_{\mu} = 1$.
Denoting $\hat  \phi_\mu(s) =\hat  \phi^a_\mu(s) \tau^a $,
we parameterize the 
3-dimensional unit vectors
$\vec \phi_\mu \equiv (\hat  \phi_\mu^1,\hat  
\phi_\mu^2,\hat  \phi_\mu^3) \in S^2$
$(\mu=1,2,3,4)$ as 
$\vec \phi_\mu =
(\sin 2 \theta_{\mu} \cos \chi_{\mu}, 
 \sin2 \theta _{\mu} \sin \chi_{\mu},\cos 2\theta_{\mu})$
using Eqs.(\ref{eq:para}) and (\ref{divided}).
The MA gauge  maximizes the third component $\hat \phi_\mu^3$
using the gauge transformation.
Under the local gauge transformation by $V(s) \equiv 1 + \{ V(s_0)-1 \} 
\delta_{ss_0} \in$ SU(2), $\hat \phi_\mu(s_0)$ is transformed as the unitary 
transformation,
\begin{eqnarray}
\hat \phi_\mu(s_0) \rightarrow \hat \phi'_\mu(s_0) \equiv 
V(s_0) \hat \phi_\mu(s_0)  V^{-1}(s_0), 
\label{eq:rot}
\end{eqnarray}
which leads to a simple rotation of the unit vectors $\vec \phi_\mu$.
In the MA gauge, 
$\displaystyle \sum_{s,\mu}$$\vec \phi_\mu$ is 
``polarized''
along the
positive third direction.
On the 4-dimension lattice with $N$ sites, 4$N$ unit vectors 
$\vec \phi_\mu(s)$ are  maximally polarized by 
$N$ gauge functions $V(s) $ in the MA gauge.
Then,  $\langle R_{\rm Abel} \rangle_{\rm MA}$ 
 is expressed as the maximal
``polarization rate'' of 4$N$ unit vectors 
$\vec \phi_\mu$ 
by suitable $N$ gauge functions $V(s)$.
On the average, this estimation of $\langle R_{\rm Abel} \rangle_{\rm MA}$
is approximately given by the estimation of the maximal 
polarization rate of 
4 unit vectors $\vec \phi_{\mu}$ by a suitable rotation with $V \in$ 
SU(2).
The lower bound of $\langle R_{\rm Abel} \rangle_{\rm MA}$ is obtained 
from the 
strong-coupling system with $\beta=0$, where link variables $U_\mu(s)$ 
are completely random. Accordingly, $\vec \phi_\mu$ can 
be regarded as 
random unit vectors on $S^2$.
The maximal ``polarization'' of 4 unit vectors $\vec \phi_\mu$
is realized by the rotation which moves  
 $\vec \phi  \equiv
$$\displaystyle \sum_{\mu=1}^4$$
\vec \phi_\mu /|
$$\displaystyle \sum_{\mu=1}^4$$
 \vec \phi_\mu|$  to the unit vector $\vec \phi^R \equiv (0,0,1)$ 
 in third direction.
Here, $\cos 2\theta_\mu^R$ after the rotation is 
identical to the inner product between $\vec \phi_\mu$ and $\vec \phi$, 
because of $\vec \phi \cdot \vec \phi_\mu = 
\vec \phi^R \cdot \vec \phi_\mu^R =
 (\hat \phi_\mu^R)^3= \cos 2\theta_\mu^R$.
Then, we estimate $\langle R_{\rm Abel} \rangle_{\rm MA} 
= \langle \cos \theta_\mu \rangle_{\rm MA}$ at $\beta=0$ as
\begin{eqnarray}
\lefteqn{ \langle \cos \theta_\mu \rangle_{\rm MA}^{\beta=0} 
\simeq  \! \{ \prod_{\mu=1}^{4} \int d U_\mu \}
\left(
\frac14 \sum_{\mu=1}^4 \cos  \theta_\mu^R 
\right) }  \nonumber \\
 &  =  & \{ \prod_{\mu=1}^{4} 
\frac{1}{\pi}
\int_0^{\frac{\pi}{2}}  d\theta_\mu \cos \theta_\mu \sin \theta_\mu 
\int_{-\pi}^{\pi}           d  \chi_\mu 
\}
\left(
\frac14 \sum_{\mu=1}^4 \cos \{ \frac12 \cos^{-1}(
\vec \phi \cdot 
\vec \phi_\mu  )  \}  \right).
\label{eq:lower}
\end{eqnarray}
Using this estimation (\ref{eq:lower}), 
we obtain  
$\langle  R_{\rm Abel} \rangle_{\rm MA} \simeq $ 0.844, 
which is close to the lattice result
$\langle R_{\rm Abel} \rangle \simeq $ 0.88  in the strong coupling 
limit ($\beta=0$).
Such a high abelian rate 
$\langle  R_{\rm Abel} \rangle_{\rm MA}$ in the MA gauge 
would provide a microscopic basis of abelian dominance for 
the infrared physics.

%
%

%
%

\section{Semi-analytical Proof of Abelian Dominance
for Confinement Force}

Abelian dominance and monopole dominance for the nonperturbative 
phenomena are numerically observed in the MA gauge in the lattice QCD 
simulations \cite{yotsuyanagi,woloshyn,giacomo2,poly,suganuma4,suganuma2,amemiya,suganuma1}.
As for confinement, monopole dominance, which means the dominant role of 
the magnetic current $k_\mu$ than that of the electric current $j_\mu$,
seems trivial if abelian dominance holds, because $j_\mu$ does not provide 
the electric confinement in 1+3 dimension. Then, as for confinement, 
abelian dominance, which means the dominant role of the diagonal element 
than that of the off-diagonal element, is the nontrivial interesting 
phenomenon observed in the MA gauge in the lattice QCD.
In this section, 
we study the origin of abelian dominance on  the string tension
as the confinement force in 
a semi-analytical manner,
considering the relation with  {\it  microscopic abelian dominance}
on the link variable\cite{ichie6}.

In the MA gauge, the {\rm diagonal element} $\cos \theta_\mu(s)$
in $M_\mu(s)$ is maximized by the gauge transformation
as large as possible. For instance, the abelian projection rate
is almost unity as 
$R_{\rm Abel}=\langle\cos \theta_\mu(s)\rangle_{\rm MA}\simeq 0.93$
at $\beta=2.4$. Then, the {\it off-diagonal element
$e^{i\chi_\mu(s)}\sin\theta_\mu(s)$ is forced to take a small value
in the MA gauge} due to the factor $\sin \theta_\mu(s)$, 
and therefore the approximate treatment
on the off-diagonal element would be allowed in the MA gauge.
Moreover, the {\it angle variable $\chi_\mu(s)$ is not
constrained} by the MA gauge-fixing condition at all,
and {\it tends to take a random value} besides the residual
${\rm U(1)}_3$ gauge degrees of freedom.
Hence, $\chi_\mu(s)$ can be regarded as a {\it random angle variable}
on the treatment of $M_\mu(s)$ in the MA gauge 
in a good approximation.

Let us consider the Wilson loop
$ \displaystyle \langle W_C[U_\mu(s)]\rangle \equiv
\langle{\rm tr}\prod_C U_\mu(s)\rangle
=\langle{\rm tr}\prod_C\{M_\mu(s)u_\mu(s)\}\rangle$
in the MA gauge.
In calculating $\langle W_C[U_\mu(s)]\rangle$,
the expectation value of $e^{i\chi_\mu(s)}$
in $M_\mu(s)$ vanishes as
\begin{eqnarray}
\langle e^{i\chi_\mu(s)}\rangle
\simeq \int_0^{2\pi} d\chi_\mu(s)\exp\{i\chi_\mu(s)\}=0,
\end{eqnarray}
when $\chi_\mu(s)$ behaves as a {\it random angle variable.}
Then, within the random-variable approximation for $\chi_\mu(s)$, 
the {\it off-diagonal factor} $M_\mu(s)$ appearing in
$\langle W_C[U_\mu(s)]\rangle$ is simply reduced as a $c$-number factor,
$
M_\mu(s) \rightarrow \cos \theta_\mu(s) \ {\bf 1},
$
and therefore the SU(2) link variable $U_\mu(s)$ in the Wilson loop
$\langle W_C[U_\mu(s)]\rangle$
is simplified as a {\it diagonal matrix,}
\begin{eqnarray}
U_\mu(s)\equiv M_\mu(s)u_\mu(s)
\rightarrow
\cos \theta_\mu(s) u_\mu(s).
\end{eqnarray}

Then, for the $I \times J$ rectangular $C$, the Wilson loop
$W_C[U_\mu(s)]$ in the MA gauge is approximated as
\begin{eqnarray}
\langle W_C[U_\mu(s)]\rangle &\equiv&
\langle{\rm tr}\prod_{i=1}^L U_{\mu_i}(s_i)\rangle
\simeq
%
%
\langle\prod_{i=1}^L \cos \theta_{\mu_i}(s_i) \cdot
{\rm tr} \prod_{j=1}^L u_{\mu_j}(s_j)\rangle_{\rm MA} \cr
&\simeq&
\langle\exp\{\sum_{i=1}^L \ln (\cos \theta_{\mu_i}(s_i))\}\rangle_{\rm MA}
\ \langle W_C[u_\mu(s)]\rangle_{\rm MA} \cr
&\simeq&
\exp\{L \langle \ln (\cos \theta_\mu(s)) \rangle_{\rm MA} \}
\ \langle W_C[u_\mu(s)]\rangle_{\rm MA},
\end{eqnarray}
where $L\equiv 2(I+J)$ denotes the perimeter length and
$ \displaystyle W_C[u_\mu(s)]\equiv {\rm tr}\prod_{i=1}^L u_{\mu_i}(s_i)$
the abelian Wilson loop.
Here, we have replaced
$ \displaystyle \sum_{i=1}^L \ln \{\cos(\theta_{\mu_i}(s_i)\}$
by its average
$L \langle \ln \{\cos \theta_\mu(s)\} \rangle_{\rm MA}$
{\it in a statistical sense}, and 
such a statistical treatment becomes more accurate for larger $I,J$ 
and becomes exact for infinite $I,J$.

In this way, we derive a simple estimation as
\begin{eqnarray}
W_C^{\rm off}\equiv
\langle W_C[U_\mu(s)]\rangle/\langle W_C[u_\mu(s)]\rangle_{\rm MA}
\simeq \exp\{L\langle \ln(\cos \theta_\mu(s))\rangle_{\rm MA}\}
\label{eq:offw}
\end{eqnarray}
for the {\it contribution of the off-diagonal
gluon element to the Wilson loop}.
From this analysis, the contribution of off-diagonal gluons 
to the Wilson loop is expected to obey the {\it perimeter law}
in the MA gauge for large loops, where the statistical
treatment would be accurate.

Now, we study the behavior of the off-diagonal contribution \\
$W_C^{\rm off}\equiv
\langle W_C[U_\mu(s)]\rangle/\langle W_C[u_\mu(s)]\rangle_{\rm MA}$
in the MA gauge using the lattice QCD, considering the theoretical 
estimation Eq.(\ref{eq:offw}).
As shown in Fig.3, we find 
that $W_C^{\rm off}$ seems to obey the
{\it perimeter law} for the Wilson loop with $I,J \ge 2$
in the MA gauge in the lattice QCD simulation with $\beta = 2.4$ and 
$16^4$.
We find also that the behavior on $W_C^{\rm off}$
as the function of $L$ is well reproduced
by the above analytical estimation with the {\it microscopic information}
on the diagonal factor $\cos\theta_\mu(s)$ as
$\langle \ln \{\cos\theta_\mu(s)\} \rangle_{\rm MA}\simeq -0.082$
for $\beta=2.4$.
Thus, the off-diagonal contribution $W_C^{\rm off}$
to the Wilson loop obeys
the perimeter law in the MA gauge, and therefore the
{\it abelian Wilson loop} $\langle W_C[u_\mu(s)]\rangle_{\rm MA}$
 should obey the
{\it area law} as well as
the SU(2) Wilson loop $W_C[U_\mu(s)]$.
From Eq.(\ref{eq:offw}),
the off-diagonal contribution to the string tension vanishes as
\begin{eqnarray}
\Delta\sigma & \equiv & 
\sigma_{\rm SU(2)}-\sigma_{\rm Abel}  \\ \nonumber 
&\equiv&
-\lim_{R,T \rightarrow \infty}
{1 \over RT}\ln \langle W_{R \times T}[U_\mu(s)]\rangle
+\lim_{R,T \rightarrow \infty}
{1 \over RT}\ln \langle W_{R \times T}[u_\mu(s)]\rangle_{\rm MA}
\cr
&\simeq&
-2 \langle \ln \{\cos\theta_\mu(s)\} \rangle_{\rm MA}
\lim_{R,T \rightarrow \infty} {R+T \over RT}=0.
\label{eq:dsig}
\end{eqnarray}
Thus, {\it abelian dominance for the string tension},
$\sigma_{\rm SU(2)}=\sigma_{\rm Abel}$,
can be proved in the MA gauge by
replacing the off-diagonal angle variable $\chi_\mu(s)$
as a random variable.

The analytical relation in Eq.(\ref{eq:offw}) indicates also that the finite
size effect on $R$ and $T$ in the Wilson loop leads to the deviation 
between the SU(2) string tension $\sigma_{\rm SU(2)}$ and the
abelian string tension $\sigma_{\rm Abel}$ as 
$\sigma_{\rm SU(2)} > \sigma_{\rm Abel}$
in the actual lattice QCD simulations.
Here, we consider this deviation 
$\Delta \sigma \equiv \sigma_{\rm SU(2)} - \sigma_{\rm Abel}$
in some detail.
Similar to the SU(2) inter-quark potential $V_{\rm SU(2)}(r)$ from
$\langle \, W_{\rm SU(2)}
 \,  \rangle \equiv
\langle \, W[U_\mu(s)] \,  \rangle$, we define the abelian inter-quark
potential $V_{\rm Abel}(r)$ and the off-diagonal contribution 
$V_{\rm off}(r)$ of the potential from
$\langle \, W_{\rm Abel} \,  \rangle \equiv
\langle \, W[u_\mu(s)] \,  \rangle$ and $W_{\rm off}$, respectively,
\begin{eqnarray}
V_{\rm SU(2)}(r) & \equiv  & -\frac{1}{Ta} {\rm ln}  \, \langle \, W_{\rm SU(2)}
({R \times T}) \,  \rangle, \nonumber \\ 
V_{\rm Abel}(r)& \equiv  & -\frac{1}{Ta} {\rm ln}  \, \langle  \,  W_{\rm Abel}
({R \times T})  \, \rangle, \nonumber \\
V_{\rm off}(r) & \equiv  & -\frac{1}{Ta} {\rm ln}   \,  W_{\rm off}
({R \times T})   
 =   -\frac{1}{Ta} {\rm ln}\,  
 \frac{ \langle \, W_{\rm SU(2)}
({R \times T}) \,  \rangle }{
 \langle  \,  W_{\rm Abel}
({R \times T})  \, \rangle } \nonumber \\
& = & V_{\rm SU(2)}(r)   - V_{\rm Abel}(r),
\end{eqnarray}
wehre $r \equiv Ra$ denotes the inter-quark distance in the physical unit.
We show in Fig.4 $V_{\rm SU(2)}(r)$,  $V_{\rm Abel}(r)$ and $V_{\rm off}(r)$
extracted from the Wilson loop with $T = 7$ in the lattice QCD simulation
with $\beta = 2.4$ and $16^4$.
As shown in Fig.4, the lattice result for $V_{\rm off}(r)$ seems to be  
reproduced by the theoretical estimation obtained from Eq.(\ref{eq:offw}),
\begin{eqnarray}
V_{\rm off}(r) = V_{\rm SU(2)}(r) - V_{\rm Abel}(r) \simeq - \frac{2(R+T)}{Ta} 
\langle \ln(\cos \theta_\mu(s))\rangle_{\rm MA}
\label{eq:voff}
\end{eqnarray}
using the microscopic information of  
$\langle \ln(\cos \theta_\mu(s))\rangle_{\rm MA} = -0.082 $ at 
$\beta = 2.4$.
From the slope of $V_{\rm off}(r)$ in Eq.(\ref{eq:voff}), we
can estimate 
$\Delta\sigma  \equiv 
\sigma_{\rm SU(2)}-\sigma_{\rm Abel}$
in the physical unit as
\begin{eqnarray}
\Delta\sigma  \equiv  \sigma_{\rm SU(2)}-\sigma_{\rm Abel}    & \simeq   &
-2 
\langle  \, \ln \, ( \, \cos \theta_\mu(s) \,  ) \,  \rangle_{\rm MA} 
\,  \frac{1}{Ta^2}
\nonumber  \\ 
& = & - 
\langle \ln( 1- \sin ^2 \theta_{\mu}(s) ) \rangle_{\rm MA}  \frac{1}{Ta^2}.
\end{eqnarray}
In the MA gauge, $ \sin ^2 \theta_\mu(s) $ takes a small value
and can be treated in a perturbation manner so that one finds
\begin{eqnarray}
\Delta \sigma
  \simeq 
\langle \sin ^2 \theta_\mu(s)  \rangle_{\rm MA}  \frac{1}{Ta^2}
= \langle (U^{1}_{\mu}(s))^2 + (U^{2}_{\mu2}(s))^2  \rangle_{\rm MA} \frac{1}{Ta^2}.
\end{eqnarray}
Near the continuum limit $a \simeq 0$, we find $U_{\mu}^a \simeq ae 
A_{\mu}^a /2$ ($a$=1,2,3) from $U_{\mu} = e^{iae A_{\mu}^a 
{\tau^a}/{2}}$, and
 then we derive the relation between $\Delta\sigma $ and the  
off-diagonal gluon in the MA gauge as
\begin{eqnarray}
\Delta\sigma 
 \simeq  
\frac{1}{4T} 
\langle (eA_{\mu}^1)^2  + (eA_{\mu}^2) ^2 \rangle_{\rm MA } =
\frac{a}{4t} \langle (eA_{\mu}^1)^2  + (eA_{\mu}^2) ^2 \rangle_{\rm MA},
\label{eq:delsig}
\end{eqnarray}
where $t \equiv Ta$ is the temporal length of the Wilson loop in the 
physical unit.
In Eq.(\ref{eq:delsig}), 
$\langle (eA_{\mu}^1)^2  + (eA_{\mu}^2) ^2 \rangle_{\rm MA }$ is the 
off-diagonal gluon-field fluctuation, and is strongly suppressed in the 
MA gauge by its definition. It would be interesting to note that
microscopic abelian dominance or the suppression of off-diagonal gluons in 
the MA gauge is directly connected to reduction of the deviation 
$\Delta\sigma$  in Eq.(\ref{eq:delsig}).
Since $\langle (eA_{\mu}^1)^2  + (eA_{\mu}^2) ^2 \rangle_{\rm MA }$ is a 
local continuum quantity, it is to be independent on both $a$ and $t$.
Hence, the deviation $\Delta\sigma $ between the SU(2) string tension 
$\sigma_{\rm SU(2)}$ and the abelian string tension $\sigma_{\rm Abel}$ 
can be removed by taking the large Wilson loop as 
$t \rightarrow \infty $ or the small mesh as $a \rightarrow 0 $ with fixed $t$.

Finally in this section, we study the origin of abelian dominance in the 
MA gauge in terms of the properties of the off-diagonal element
\begin{eqnarray}
c_\mu(s) \equiv e^{i \chi_{\mu}(s)} \sin \theta_{\mu}(s)
\end{eqnarray}
of $M_{\mu}(s)$ in the link variable $U_\mu(s)$, considering the validity 
of the random-variable approximation for $\chi_\mu(s)$ in the MA gauge.
{\it In the above  treatment, the contribution of the off-diagonal element in 
the link variable $U_{\mu}(s)$ is completely dropped off, and its 
effect indirectly  remains as the appearance of the $c$-number
factor $\cos  \theta_{\mu}(s)$ in the link variable. Such a reduction 
of the contribution of the off-diagonal elements is brought by the two 
relevant features on the two local variables, $\theta_{\mu}(s)$
and  $\chi_{\mu}(s)$, in the MA gauge.
One is the microscopic abelian dominance as 
$\langle \cos \theta_{\mu}(s) \rangle_{\rm MA} \simeq 1$ in the MA 
gauge, and 
the other is the randomness of the off-diagonal variable $\chi_{\mu}(s)$.  }
\begin{enumerate}
\item In the MA gauge, the microscopic abelian dominance holds as 
$\langle \cos \theta_{\mu}(s) \rangle_{\rm MA}$ $\simeq 1$, and the absolute value of the 
off-diagonal element 
$|c_{\mu}(s)| = |\sin  \theta_{\mu}(s) |$ is strongly reduced. Such a 
tendency becomes more significant as $\beta$ increases.
\item The off-diagonal angle variable  $\chi_{\mu}(s)$ is not constrained by the 
MA gauge-fixing condition at all, and tends to be a random variable.
In fact, $\chi_{\mu}(s)$ is affected only by the action factor $e^{- 
\beta S_{\rm QCD}}$ in the QCD generating functional, but the effect 
of the action to $\chi_{\mu}(s)$ is quite weaken due to the small factor 
$\sin  \theta_{\mu}(s)$ in the MA gauge.
The randomness of  $\chi_{\mu}(s)$ tends to vanish the contribution of 
the off-diagonal elements.
\end{enumerate}

Here, the randomness of the off-diagonal angle-variable
$\chi_\mu(s)$ is closely related to the 
microscopic abelian dominance.
In fact, the randomness of $\chi_{\mu}(s)$ is controlled 
only by the action factor $e^{- 
\beta S_{\rm QCD}}$ in the QCD generating functional, however the effect 
of the action to $\chi_{\mu}(s)$ is quite weaken due to the small factor 
$\sin  \theta_{\mu}(s)$ in the MA gauge, because $\chi_{\mu}(s)$ always 
accompanies $\sin \theta_{\mu}(s)$  in the link variable $U_{\mu}(s)$.
Near the strong-coupling limit $\beta \simeq 0$,
the action factor $e^{-\beta S_{\rm QCD}}$ brings almost no 
constraint on  $\chi_{\mu}(s)$ in the MA gauge.
The independence of $\chi_{\mu}(s)$ from the action factor is enhanced 
by the small factor $\sin \theta_{\mu}(s)$
accompanying $\chi_{\mu}(s)$.
Hence, 
 $\chi_\mu(s)$ behaves as a random angle-variable almost exactly, and
 the contribution of the off-diagonal element is expected 
to disappear in the strong-coupling region.
As $\beta$ increases, the action factor 
$e^{-\beta S_{\rm QCD}}$ becomes relevant and will reduce the 
randomness of 
$\chi_{\mu}(s)$ to some extent.
Near the continuum limit $\beta \rightarrow \infty$, however, the 
factor $\sin \theta_{\mu}(s)$ tends to approach 0 in the MA gauge as shown in 
Fig.2(b), and hence such a constraint on $\chi_{\mu}(s)$ from the 
action is largely reduced, and the strong  randomness of
$\chi_{\mu}(s)$ is expected to hold there.
Moreover, the reduction of the absolute value $| c_{\mu}(s)| = | \sin 
\theta_{\mu}(s)|$ itself
further reduces the 
contribution of the off-diagonal element $|c_{\mu}(s)|$ in the MA gauge.
 
Now, we examine the randomness of $\chi_\mu(s)$
using the lattice QCD simulation.
 It should be noted  that
the residual U(1)$_3$ gauge degrees of freedom
should be fixed to extract  $\chi_\mu(s)$ itself,
because $\chi_\mu(s)$ is 
the U(1)$_3$ gauge variant.
To this end, we add the U(1)$_3$ lattice Landau
gauge \cite{amemiya}, which maximizes
\begin{eqnarray}
R \, = \, \sum_{s, \mu} Re \, {\rm tr} \, u_\mu(s)
\end{eqnarray}
using the residual U(1)$_3$ gauge transformation in the MA gauge.
In the U(1)$_3$ Landau gauge, there remains no local symmetry and 
the lattice variable becomes mostly continuous
and approaches to the continuum field under the constraint of the MA gauge fixing.
For the test of the randomness of $\chi_{\mu}(s)$, we calculate the 
probability distributions of $\chi_{\mu}(s)$ and the correlation 
between $\chi_\mu(s)$ and $\chi_\mu(s+ \hat \nu)$
in the MA gauge with the U(1)$_3$-Landau gauge.
If  $\chi_{\mu}(s)$ is a random angle variable, there is no bias on 
the distribution of $\chi_{\mu}(s)$ and no correlation is 
observed between $\chi_\mu(s)$ and $\chi_\mu(s+ \hat \nu)$.
We show in Fig.5(a) the probability distributions $P(\chi_{\mu})$
and $P(\theta^3_{\mu})$ at $\beta=2.4$.
Unlike $P(\theta^3_{\mu})$, $P(\chi_{\mu})$ is flat distribution
without any structure in the whole region of $\beta$,
which is necessary condition of the random angle variable.
We show in Fig.5(b)
the probability distribution $P(\Delta \chi)$
of the correlation
\begin{eqnarray}
\Delta \chi(s) \equiv d(\chi_\mu(s),\chi_\mu(s+ \hat \nu)) \equiv 
{\rm mod}_{\pi}|\chi_\mu(s)-\chi_\mu(s+ \hat \nu)| \in [0,\pi],
\end{eqnarray}
which is the difference between two neighboring angle variables,
at $\beta$=0, 1.0, 2.4, 3.0.
In the strong-coupling limit $\beta=0$,
$\chi_\mu(s)$ is a completely random variable, and
there is no correlation between neighboring $\chi_\mu$.
In the strong-coupling region as $\beta \le 1.0$,
almost no correlation is observed between neighboring $\chi_\mu$,
which suggests the strong randomness of $\chi_\mu(s)$.
As a remarkable feature, the correlation between
neighboring $\chi_\mu$ seems weak
even in the weak-coupling region as $\beta \ge 2.4$,
where the action factor $e^{-\beta S_{\rm QCD}}$ becomes dominant
and remaining variables $\theta_\mu^3(s)$ and $\theta_\mu(s)$ behave as 
continuous variables with small difference between their neighbors as
$\Delta \theta_\mu^3 \simeq 0$ and $\Delta \theta_\mu \simeq 0$.
Such a weak correlation of neighboring $\chi_\mu$ would be originated 
from the reduction of the accompanying factor $\sin \theta_{\mu}(s)$ in the MA gauge.
Moreover, in the weak-coupling region,
the smallness of $\sin \theta_\mu(s)$ makes $c_\mu(s)$
more irrelevant in the MA gauge, which permits some approximation on $\chi_\mu(s)$.
Thus, the random-variable approximation for $\chi_\mu(s)$
would provide a good approximation
in the whole region of $\beta$ in the MA gauge.
To conclude, the origin of abelian dominance for confinement in the MA 
gauge is stemming from the strong randomness of the off-diagonal angle 
variable $\chi_\mu(s)$ and the strong reduction of the off-diagonal 
amplitude $|\sin \theta_\mu(s)|$ as the result of the MA gauge fixing.

\section{Summary and Concluding Remarks}

In the 't~Hooft abelian gauge, QCD is reduced into
an abelian gauge theory, and the color-magnetic
monopole appears as the topological object
in the constrained nonabelian gauge manifold
corresponding to the nontrivial homotopy group
$\Pi_2(SU(N_c)/U(1)^{N_c-1})=Z^{N_c-1}_\infty$.
Hence, if off-diagonal gluons can be neglected
and the monopole is condensed,
the QCD vacuum in the abelian gauge 
is described as the abelian dual superconductor 
and the confinement mechanism is understood
as the dual Meissner effect.

In relation with the dual Higgs picture 
for the confinement mechanism in QCD, 
we have studied the mathematical features of 
the abelian gauge fixing, 
the local gluon properties 
in the maximally abelian (MA) gauge, 
and the origin of abelian dominance for confinement 
in a semi-analytical manner with the help of the lattice QCD.

First, we have studied the residual symmetry in the abelian gauge,
with paying attention to the global Weyl symmetry, which
can remain as the relic of SU($N_c$).
The global Weyl symmetry provides the ambiguity on 
the electric and magnetic charges, and 
persistently remains in the MA gauge.
Considering the abelian gauge fixing in terms of the coset space of 
the fixed gauge symmetry, 
we have derived the criterion on the SU($N_c$)-gauge invariance 
of the operator in the abelian gauge: 
if the operator defined in the abelian gauge 
is invariant under the residual gauge transformation, 
it is also SU($N_c$)-gauge invariant.

Second, in the continuum SU($N_c$) QCD,
we have expressed the MA gauge fixing 
using the SU($N_c$)-covariant derivative operator.
The local MA-gauge fixing condition 
and the composite Higgs field $\Phi[A_\mu(s)]$ to be
diagonalized are naturally derived from this expression of the MA gauge.

Third, we have examined the abelian projection rate $R_{\rm Abel}$, 
the overlapping factor between SU(2) and abelian link variables,
in the lattice formalism.
In the MA gauge, we have found 
the high abelian projection rate as 
$\langle R_{\rm Abel}\rangle_{\rm MA} \simeq 1$ 
for the whole region of $\beta$, which means 
microscopic abelian dominance on the link variable.
Using the statistical consideration, 
we have analytically estimated the lower bound of the 
abelian projection rate in the MA gauge as 
$\langle R_{\rm Abel}\rangle_{\rm MA} \ge 0.84$, 
which seems consistent with the lattice result 
$\langle R_{\rm Abel}\rangle_{\rm MA} \ge 0.88$.

Finally, we have studied abelian dominance in terms of 
off-diagonal gluons in the Wilson loop in the MA gauge.
In the SU(2) link variable, the off-diagonal angle variable
is not constrained by the MA-gauge fixing condition at all,
and tends to take random values 
besides the residual gauge degrees of freedom.
By approximating the off-diagonal angle variable as a random variable,
we have proved that 
the contribution of off-diagonal gluons to the Wilson loop,
$W_{\rm off}\equiv \langle W_{\rm SU(2)} \rangle/
\langle W_{\rm Abel}\rangle_{\rm MA}$, 
obeys the perimeter law in the MA gauge,
which is numerically confirmed 
using the lattice QCD Monte Carlo simulation.
Thus, we have showed exact abelian dominance for the string tension 
as $\sigma_{\rm SU(2)}=\sigma_{\rm Abel}$, 
although the finite size effect of the Wilson loop 
in the actual lattice QCD simulation 
leads to small deviation as $\sigma_{\rm SU(2)} > \sigma_{\rm Abel}$.

In conclusion, we have found that the origin of abelian dominance for 
confinement in the MA gauge
is stemming from the strong randomness of the off-diagonal angle 
variable $\chi_\mu(s)$ and the strong reduction of the off-diagonal 
amplitude $|\sin \theta_\mu(s)|$, and these two remarkable features on 
the local variables $\chi_\mu(s)$ and  $\theta_\mu(s)$ are peculiar to the MA 
gauge fixing.

One of authors (H.I.) is supported by 
Research Fellowships of the Japan Society for the 
Promotion of Science for Young Scientists.
One of authors (H.S.) is supported in part by 
Grant for Scientific Research (No.09640359) from 
the Ministry of Education, Science and Culture, Japan.
The lattice QCD simulations have been performed on VPP500 at RIKEN and
sx4 at RCNP.

\newpage

\section*{Figure Captions}

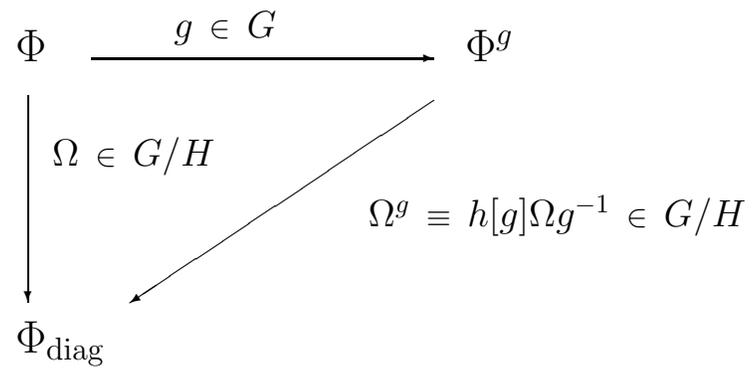
Fig.1: The gauge transformation property of $\Phi$ and
gauge function $\Omega \in$ $G/H$.
The abelian gauge fixing is realized by $\Omega$ $\in$ $G/H$ so as to
diagonalize $\Phi$.
(a) After the gauge transformation by $^\forall g\in G$,  
the operator $\Phi^g$ is diagonalized by the gauge function   $\Omega 
^g=h[g]\Omega g^\dagger$
$\in$ $G/H$.
(b) The gauge transformation property of the operator $O^\Omega$
defined in the abelian gauge.
If $O^\Omega$ is $H$-invariant,
$O^\Omega$ is 
found to be invariant under the whole gauge transformation of $G$. \\

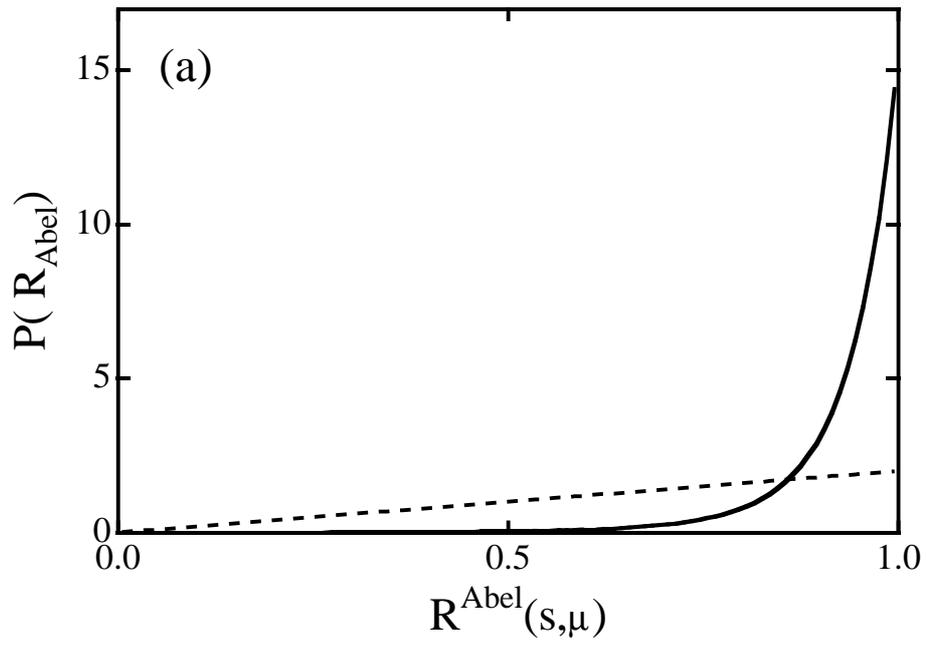
Fig.2: 
(a) The probability distribution $P(R_{\rm Abel})$
of the abelian projection rate $R_{\rm Abel} $
at $\beta =2.4$ on the $16^4$ lattice from 40 gauge configurations.
The solid curve denotes $P(R_{\rm Abel})$ in the MA gauge,
and  the dashed line denotes $P(R_{\rm Abel})$
without gauge fixing.
(b) The average of the abelian projection rate $\langle  R_{\rm Abel} 
\rangle$ in the MA gauge
as the function of $\beta$.
For comparison, we plot also $\langle  R_{\rm Abel} \rangle$ without gauge 
fixing. \\

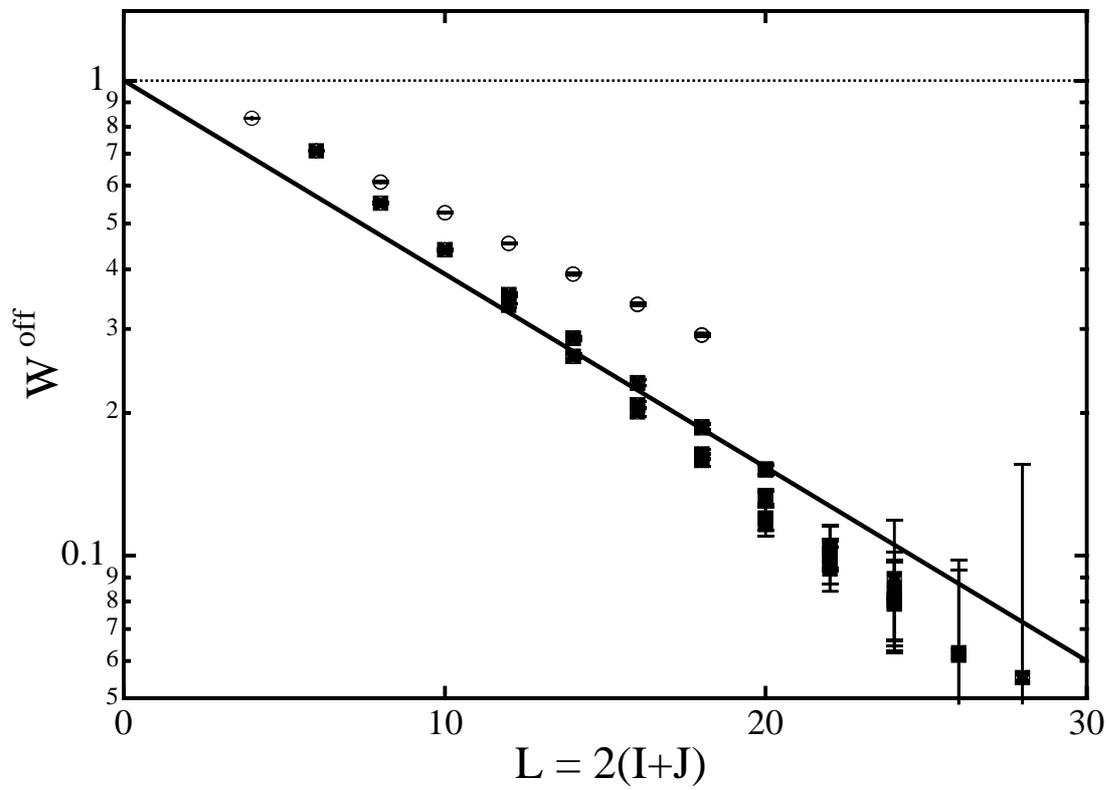
Fig.3: 
The off-diagonal gluon contribution on the Wilson loop,
$W^{\rm off} 
 \equiv  \frac{  \langle  W_C[U_\mu(s)]  \rangle }
                          {  \langle  W_C[u_\mu(s)]  \rangle }$,
as the function of the perimeter length $L\equiv 2 (I + J)$
in the MA gauge on $16^4$ lattice with $\beta = 2.4$.
The thick line denotes the theoretical estimation 
in Eq.(\ref{eq:offw})
with the microscopic input 
$\langle \ln \{\cos \theta_\mu(s)\}\rangle_{\rm MA} \simeq -0.082$ 
at $\beta=2.4$.
The data of the Wilson loop with $I=1$ or $J=1$ are distinguished by 
the circle. \\

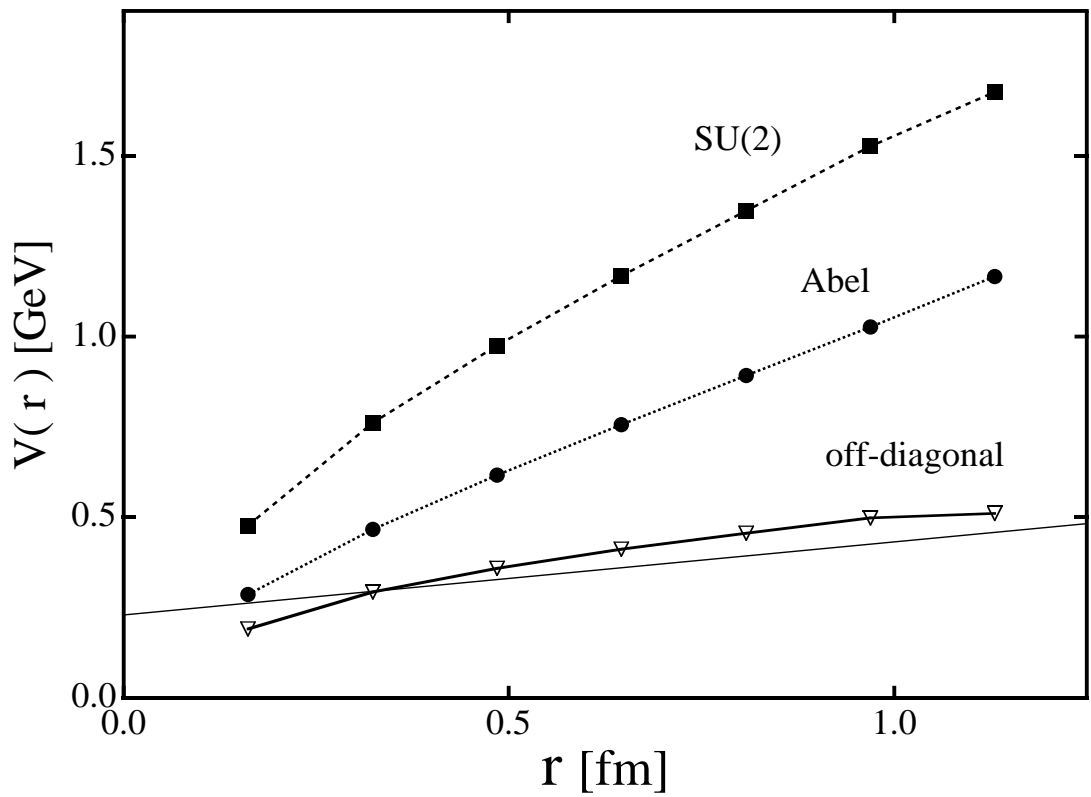
Fig.4: 
The inter-quark potential $V(r)$ as the function of
the inter-quark distance $r$.
The lattice data are obtained from the Wilson loop in the MA gauge
on $16^4$ lattice with $\beta = 2.4$ and $T=7$.   
The square, the circle and the rhombus denote the full SU(2), 
the abelian
and the off-diagonal contribution of the static potential,
respectively.
The thin line denotes the theoretical estimation
in Eq.(\ref{eq:voff}).
Here, the lattice spacing $a$ is determined so as to produce $\sigma = 0.89 $ GeV/fm.
Due to the artificial finite-size effect of the Wilson loop,
the off-diagonal contribution $V^{\rm off}$ gets a slight linear part. \\

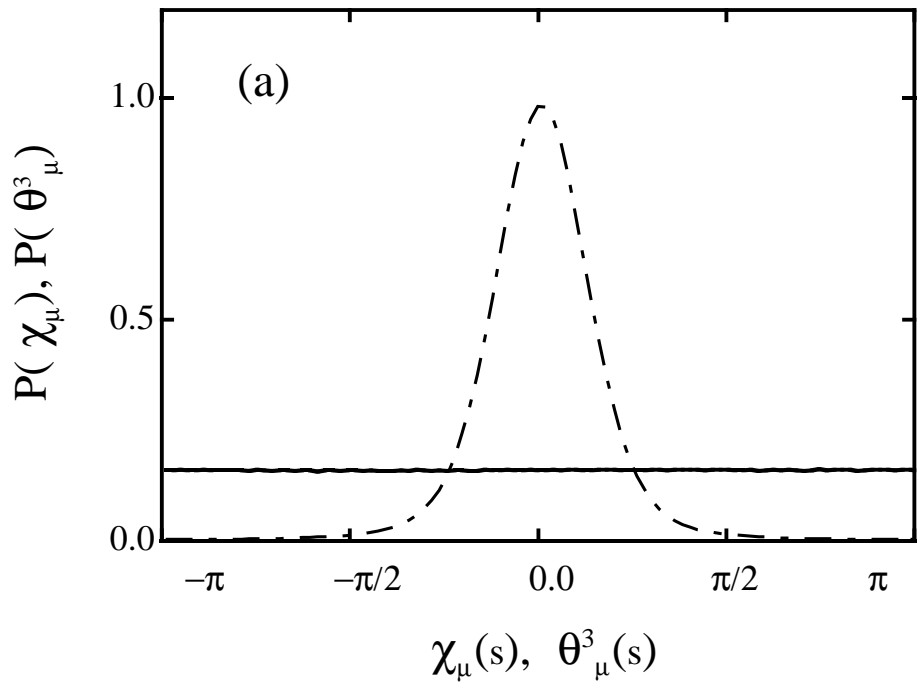
Fig.5: 
(a) The probability distributions $P(\chi_\mu)$ (solid line)
and $P(\theta^3_\mu)$ (dash-dotted curve)
in the MA gauge with the U$(1)_{3}$ Landau gauge
at $\beta =2.4$ on the $16^4$ lattice from 40 gauge configurations.
(b) The probability distribution $P(\Delta \chi)$
of the correlation $\Delta \chi \equiv {\rm 
mod}_{\pi}(|\chi_\mu(s)-\chi_\mu(s+ \hat \nu)|)$ in the same gauge 
at $\beta$ = 0 (thin line), 1.0 (dotted curve), 2.4 (solid curve), 
3.0 (dashed curve).

\newpage

\begin{figure}[bt]


{\Large Fig.1}

\vspace{4cm}
\centering \leavevmode
{\tt    \setlength{\unitlength}{0.92pt}
\begin{picture}(214,352)
\thinlines    \put(70,22){{{\large $h[g]$} $\in$ {\large $H$}}}
              \put(20,81){{\large $\Omega $}}
              \put(70,150){{{\large $g$} $\in$ {\large $G$}}}
              \put(150,271){{{\large $\Omega^g $} $ \equiv $ {\large $h[g] 
\Omega 
              g^{-1}$} $\in$ {\large $G/H$} }}
              \put(20,296){{{\large $\Omega$} $\in$ {\large 
              $G/H$}}}
              \put(70,349){{{\large $g$} $\in$ {\large $G$}}}
              \put(210,82){{\large $\Omega^g $} $\equiv$ {\large $h[g] 
\Omega 
              g^{-1}$}}
              \put(5,9){{\large $O^\Omega$}}
              \put(190,9){{\Large $(O^{\Omega})^{h[g]}$}}
              \put(190,141){{\Large $O^g$ }}
              \put(5,141){{\Large $O$ }}
              \put(5,219){{\Large $\Phi_{\rm diag}$}}
              \put(190,339){{\Large $\Phi^g$}}
              \put(5,339){{\Large $\Phi $}}
              \put(36,12){\vector(1,0){141}}
              \put(196,126){\vector(0,-1){95}}
              \put(10,125){\vector(0,-1){95}}
              \put(36,141){\vector(1,0){141}}
              \put(10,216){{}}
              \put(10,221){{}}
              \put(177,323){\vector(-3,-2){125}}
              \put(10,325){\vector(0,-1){85}}
              \put(36,340){\vector(1,0){141}}
\end{picture}}
\label{gfig2}
\vspace{0cm}
\end{figure}

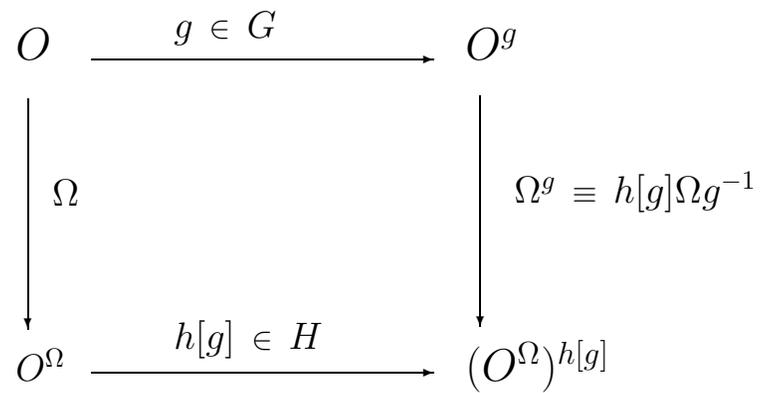
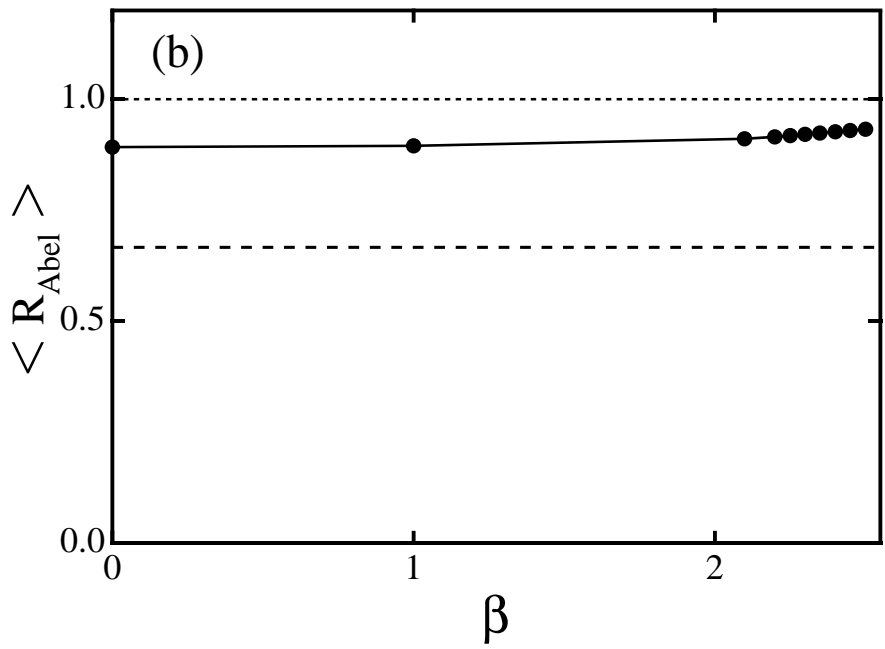
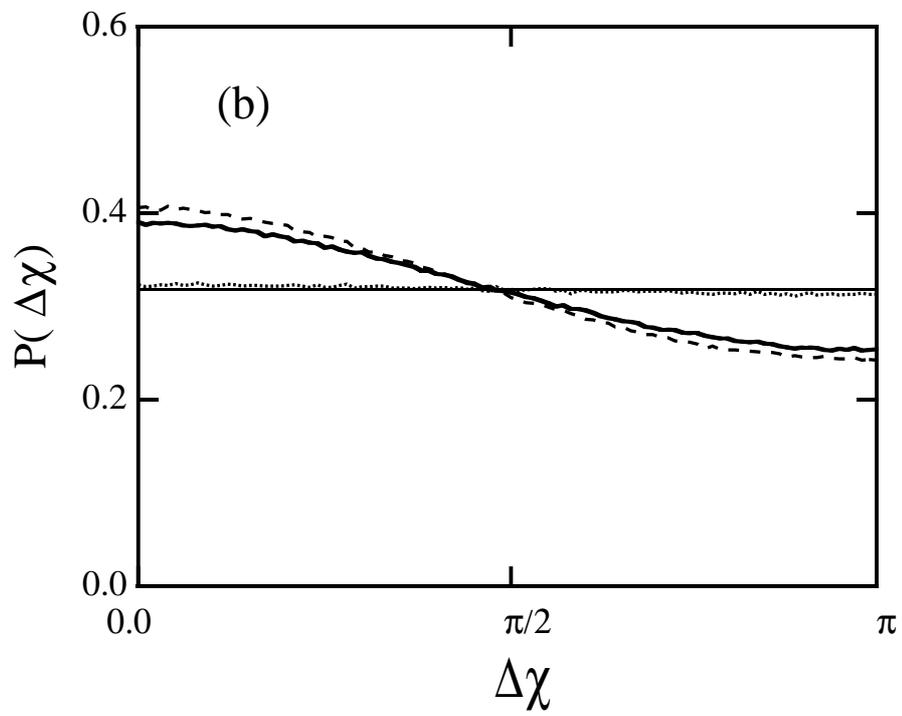
\end{document}